\documentstyle[epsf,epsfig,amssymb]{article}
\newcommand{\ppbar}{$p\overline{p}\ $}
\newcommand{\ttbar}{$t\overline{t}\ $}
\newcommand{\ebar} {\hbox{E\kern-0.5em\lower-0.1ex\hbox{/}}}
\def\Journal#1#2#3#4{{#1} {\bf #2}, #3 (#4)}

\def\NPB{{\em Nucl.\ Phys.} B}
\def\PLB{{\em Phys.\ Lett.} B}
\def\PRL{\em Phys.~Rev.~Lett.}
\def\PRD{{\em Phys.~Rev.} D}

\begin{document}

\centerline{~~~}
{\rightline{\Large 
%CDF/PUB/CDF/PUBLIC/5162
}}
{\rightline{\Large November 22, 1999}}
%\vspace*{2cm}
\vspace*{0.5cm}

%                          Title
\begin{center}
{\Large \bf Recent Results from CDF
\footnote{Proceedings of the XIV International Workshop on High--Energy Physics and Quantum Field Theory 
(QFTHEP '99), Moscow, Russia, May 1999.}
} \\

\vspace{4mm}

%                      author/address
Michele Gallinaro \\
The Rockefeller University \\
1230 York Ave., New York, NY 10021, USA\\
michgall@fnal.gov\\
\end{center}

%                       Abstract
\begin{abstract}
We present the latest results from the CDF experiment at the Tevatron Collider
in \ppbar collisions at $\sqrt{s} = 1.8$~TeV.
The large data sample collected during Run 1, from 1992 until 1995,
allows measurements in many domains of high--energy physics. 
Here, we report
on the first measurement of $\sin(2\beta)$, a CP violation parameter, and on
an improved measurement of the top quark cross section.
We also report on searches for the so--far elusive Higgs boson,
and for SUSY, through searches for direct production of top
and bottom scalar quarks.
Finally, we outline the prospects for the physics during the upcoming Run 2,
ready to start in the upcoming year 2000.
\end{abstract}

\section{Introduction}
At the Tevatron Collider bunches of protons and antiprotons are accelerated and
collided at an energy of $\sqrt{s} = 1.8$~TeV. The Collider Detector at Fermilab
(CDF) and D\O ~ detectors have collected large samples of data. Here, we
present some of the latest results from the CDF experiment.
Data were collected during two periods: Run 1A groups about 20~pb$^{-1}$ and Run
1B about 90~pb$^{-1}$, for a total period of time that goes from 1992 until the end of 1995.

The CDF detector~\cite{CDF} relies on a high resolution tracking system with a central
tracking drift chamber immersed in a 1.4 Tesla magnetic field. In
addition, a high precision silicon micro-vertex detector was added inside the
drift chamber and near the beam--pipe to allow for
% excellent and precise measurements of short lived particles and 
precise vertex determination. 
Outside the tracking system, electromagnetic and
hadronic calorimeters provide excellent measurement of particle energy. 
Muon chambers located outside the magnetic yoke complete the tracking system.

The results presented here include the direct measurement of the 
CP violation parameter of $\sin(2\beta)$.
Recently, the CDF collaboration also improved the measurement 
of the \ttbar production cross section, and continues the investigation for new physics, 
including Higgs boson and SUSY searches.
We conclude the paper and highlight the improved capabilities in Run II, and
summarize the upgrade projects under construction for the CDF detector.

\section{CP Violation: measurement of \boldmath{$\sin(2\beta)$}}
The origin of CP violation has been an outstanding problem since its
first discovery in $K^0_L\rightarrow\pi^+\pi^-$ decays. After 35 years of the
discovery of CP violation, the kaon system was still the only place where CP
violation had been observed. 
However, neutral B mesons are expected to yield large CP violating effects.
Here, we report on a direct measurement that indicates a CP violation effect in
$B^0\rightarrow J/\Psi K^0_s$ decays.

CP violation can be explained in an elegant way in the Standard Model (SM) with
three generations, as suggested by Cabibbo, Kobayashi and Maskawa~\cite{CKM}.
They proposed that quark mixing is the cause. The CKM matrix relates the mass
and weak eigenstates of the quarks.
With three generations of quarks, the CKM matrix has a physical
complex phase capable of accommodating CP violation.

Similar to the kaon system, CP violating effects are expected in $B$
mesons.
The effects of CP violation in the inclusive decay channels of the $B$ mesons 
are still too small ($\sim 10^{-3}$) to be detected~\cite{CLEO}. 
Yet, the interference due to mixing of
$B^0_d$ decays to the same CP state could show large violations.
CP violation can manifest itself as an asymmetry in the decay rate of 
particle versus anti-particle to a particular CP eigenstate.
For example, a $B^0_d$ may decay directly to $J/\Psi K^0_s$, or it may oscillate to a
$\overline{B^0_d}$ and then decay to $J/\Psi K^0_s$. The two paths have a phase
difference and the interference results in an asymmetry:

$${\cal A}_{CP}(t) = {\overline{B^0_d}(t) - B^0_d(t)\over
		    \overline{B^0_d}(t) + B^0_d(t) } = sin(2\beta)\cdot sin(\Delta
		    m_d t),$$
\noindent
where $B^0_d(t)$  $[\overline{B^0_d}(t)]$ is the number of $J/\Psi K^0_s$ decays at
proper time $t$ from mesons produced as $B^0_d$ $[\overline{B^0_d}]$, and 
$sin(\Delta m_d t)$ is the mixing oscillation due to the decay interference.
In the SM, the CP asymmetry is proportional to $\sin 2\beta$,
where $\beta$ is an angle of the unitarity triangle. In order to measure the
asymmetry ${\cal A}_{CP}$ we have to identify the flavor of the B meson at the
time of production. Unfortunately, the tagging algorithms are not perfect, and
the ``true'' asymmetry is ``diluted'' by a dilution factor ($D$) that takes into
account the ``mistakes'' of the tagging algorithms, which incorrectly identify 
the flavor.
The dilution factor is defined as $D=2P-1$, where $P$ is the probability to have a correct tag.
Therefore, the observed asymmetry
${\cal A}^{obs}_{CP}= D \cdot {\cal A}_{CP}$, is reduced by the dilution parameter.

We take advantage of the large $B$ meson production cross section at the
Tevatron, and use a sample of $J/\Psi K^0_s$ decays to measure $\sin 2\beta$.
This measurement uses a data sample which is larger by a factor of two with respect to
our previous measurement \cite{old_sin2beta} by including $J/\Psi K^0_s$ 
events not reconstructed in the SVX.
Here, in addition to the tagging algorithm which we previously used in 
Ref.~\cite{old_sin2beta}, we also employ two additional tagging algorithms which 
further enlarge the effective statistical power our sample size.
We reconstruct $B\rightarrow J/\Psi K^0_s$ events similarly to previous
analyses at CDF. Then, we categorize the events according to the tagging
algorithm used. We measure dilutions and efficiencies for each class of events and use
a maximum likelihood fitting procedure to extract the result for $\sin 2\beta$.

\begin{figure}[thb]
\begin{minipage}[]{.46\linewidth}
\centerline{\psfig{figure=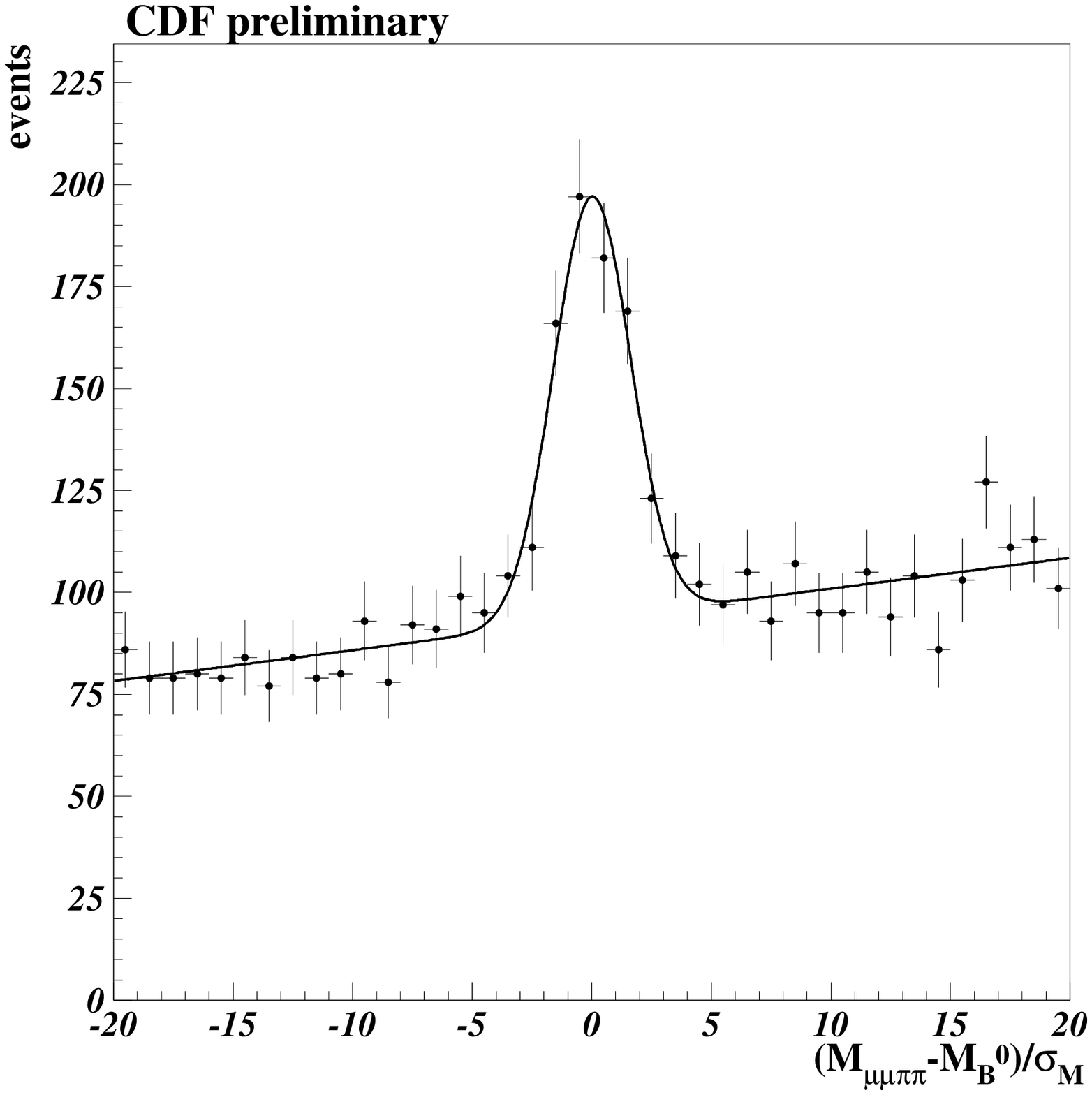,height=8.5cm,width=8.5cm}}
\caption{\label{sin2beta_sample} 
Mass distribution of the $J/\Psi K^0_s$ candidates. The fit of a 
gaussian distribution plus a linear background fit agrees 
well with the data. 
The invariant mass distribution is normalized to the world average $B^\circ_d$
mass ($M_{B^\circ}$), and to the uncertainty in the fit ($\sigma_M\sim 10$~MeV/c$^2$).}
\end{minipage}
\hspace*{0.5cm}
\begin{minipage}[]{.46\linewidth}
\centerline{\psfig{figure=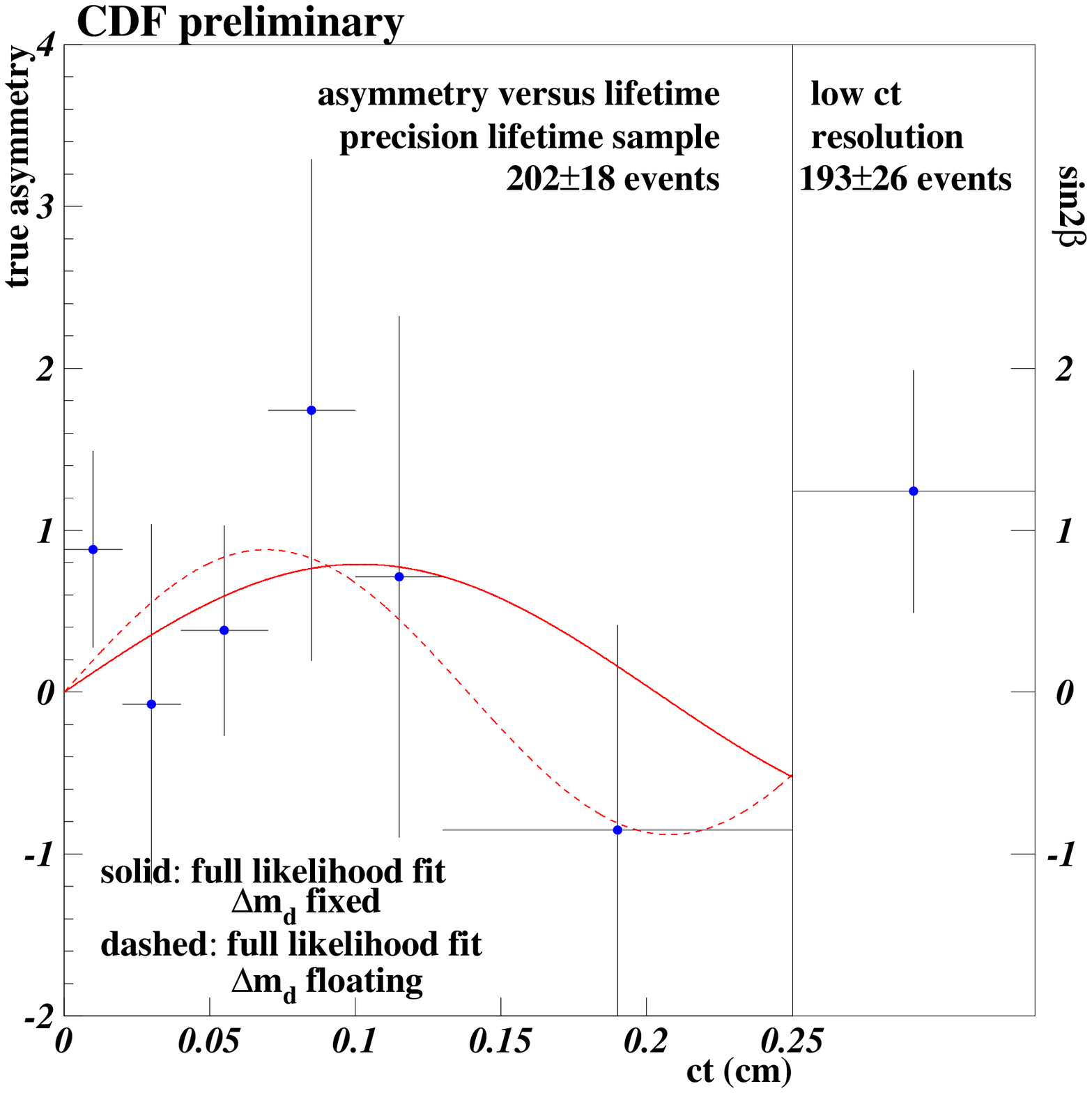,height=8.5cm,width=8.5cm}}
\caption{\label{sin2beta_result} 
The true asymmetry as a function of lifetime for $B\rightarrow J/\Psi K^0_s$ events.
The SVX data is shown in proper time bins on the left, and a single bin for
non--SVX data on the right.}
\end{minipage}
\end{figure}

We reconstruct $J/\Psi$ events decaying to $\mu^+\mu^-$. We identify $K^0_s$ candidates
by reconstructing oppositely charged tracks to the 
$K^0_s\rightarrow\pi^+\pi^-$ hypothesis. 
We constrain the $\mu^+\mu^-$ and $\pi^+\pi^-$ pairs to the world average masses.
The reconstructed invariant mass of the $\mu^+\mu^-\pi^+\pi^-$ system
for all the candidates is shown in Figure~\ref{sin2beta_sample}.
About one half (202$\pm$18 events as opposed to 193$\pm$26 events)
of the final $B^\circ$ signal sample contains a precise lifetime information
as both muons are contained in the silicon vertex detector (SVX).
The precise lifetime information allows us to make a time--dependent measurement
of the oscillation.
We can also use the events that are not contained within the SVX acceptance.
In this case, CP asymmetry is still present in the form of a time--integrated
measurement. The statistical power is only reduced by a factor of one third in
our non--SVX sample, where the lifetime information is lost.

The measurement of ${\cal A_{CP}}$ is possible if we can determine the flavor
($B$ or ${\overline B}$) of the $B$ meson at the time of production. A tagging
method with dilution $D$ and tagging efficiency $\epsilon$ yields an uncertainty on 
$\sin 2\beta$ which scales as 1/$\sqrt{\epsilon D^2 N}$, with $N$ events. Therefore,
$\epsilon D^2$ measures the effective tagging power.
Here, we use a combination of different tagging methods.
We use three taggers.
We use a Same Side Tagger (SST), which relies on selecting a track (supposedly
from a $\pi^{\pm}$) near the $B$
meson, and it is correlated to the $b$--flavor~\cite{sst}.
The SST tag is a candidate track with a
$p_T>400$ MeV/c, reconstructed in the SVX. The SST tagger was used in our
earlier measurement of $\sin 2\beta$. The SST dilution is measured to be
$(16.6\pm 2.2)\%$ in the $J/\Psi K^0_s$ sample, for events reconstructed in the SVX.
Here, we extend the SST algorithm to events outside the SVX coverage (we drop
the impact parameter cut), and we find a dilution factor of $(17.4\pm 3.6)\%$.

In order to further increase the effective number of candidate events, we use 
two additional taggers that determine the flavor of the ``other'' $B$
meson; we can also refer to them as ``opposite side taggers''.
We use the Soft Lepton Tagger (SLT) which
determines the charge of the lepton from the $b\rightarrow l$ decay~\cite{slt}.
Lepton tracks have a $p_T>1(2)$~GeV/c$^2$ for electrons (muons).
The dilution is measured from $B^+\rightarrow J/\Psi K^+$ sample and we 
find $D=62.5\pm 14.6\%$.
The other opposite side tagger is a Jet Charge (JETQ) tagger~\cite{jetq}. The flavor is
determined by a measure of the average of the charge of the opposite side
jet. The jet is formed by an algorithm that combines tracks ($p_T>0.4$~GeV/c) around a seed
track with $p_T>1.75$~GeV/c, up to the mass of the $B$. The dilution again is measured from
$B^+\rightarrow J/\Psi K^+$ sample, and is $23.5\pm 6.9\%$.

In general, the ideal tagger should have high dilution and high efficiency.
In reality, the SLT tagger has high dilution but low efficiency, while
both SST and JETQ taggers have higher efficiency and low dilution values.
By coincidence, all of the three taggers measured $\epsilon D^2\sim 2\%$.
Therefore, we have effectively tripled the size of our event sample with
respect to ~\cite{old_sin2beta}.
The combined $\epsilon D^2$ for the three taggers is $6.3\pm1.7\%$. 
In conclusion, our sample of 400 events
corresponds to $\sim 25$ perfectly tagged $J/\Psi K^0_s$ events.

We use a negative log--likelihood method to determine the best value 
for $\sin 2\beta$. The $B^0$ lifetime and the $\Delta m_d$ values are fixed to
the world average values (1.54$\pm$0.04 psec and 0.464$\pm$0.018 psec$^{-1}$).
The fit yields a value $\sin 2 \beta = 0.79 ^{+0.41}_{-0.44}$ (stat. +
syst.) for all the taggers combined~\cite{new_sin2beta}.
In Figure~\ref{sin2beta_result} we show the distributions of the asymmetry 
both for the SVX and for the non--SVX samples.
The precision of our direct measurement of $\sin 2 \beta$ is still
dominated by the statistical uncertainty ($\pm$0.36). 
This measurement provides for the
first time an indication of CP violation in the $B$ system, and it is
consistent with the SM expectations.

Definite observation of CP violation in the $B$ mesons requires a better precision.
In Run II, with an integrated luminosity of $\sim 2$~fb$^{-1}$, we expect to reconstruct 
$\sim 10^4 J/\Psi K^0_s$ from dimuons, which will give a total uncertainty of about
$\pm 0.08$. Triggering on $J/\Psi\rightarrow e^+ e^-$ may increase the sample
by 50\%. CDF is also upgrading the detector with a Time--of--Flight system
which will help flavor identification.

\section{Top Quark Cross Section}
At the Tevatron top quarks are produced mainly in pairs, via $q\bar{q}$
annihilation, with a theoretically expected cross section of only 
about 5~pb ~\cite{xsec_theory}. 
After several years of data taking during Run I, only about 600 \ttbar 
events were produced. When we fold in the geometrical acceptances, identification
criteria and selection requirements, the data set becomes much smaller.

In the framework of the SM each top quark decays into a $W$ and a $b$
quark. The final state of a \ttbar decay therefore has two $W$ bosons and two
$b$ quarks. Direct evidence of the existence of the top quark was presented by
CDF ~\cite{top_evidence} in early 1994. Later, in 1995, both the CDF
and the D\O~ collaborations announced conclusive discovery of the
top quark~\cite{top_observation}.
A key to the identification of \ttbar events has proved to be the ability to
{\it tag} the $b$ quarks. Tagging of $b$ quarks can be
achieved at CDF using the silicon micro-vertex detector (SVX), and we can attain a
$b$ tagging efficiency of about $\sim 50\%$ in a \ttbar event, by finding a
secondary vertex from the interaction point (SVX tagging). Also, we can
identify $b$ quarks by means of a lepton ($e$ or $\mu$) 
from $b$ decays, with an efficiency of $\sim 20\%$. This method is 
referred to as Soft Lepton Tagger (SLT).
We measure the \ttbar production cross section in several decay channels and all the
channels yield consistent results. A final result is obtained by combining the
cross sections in the individual channels.

In this paper we report on an improved measurement of the \ttbar cross section
in the ``lepton+jets'' channel, where one $W$ decays leptonically and the other
$W$ decays hadronically. Here, we identify the $b$ quarks through the SVX tagging method.
This yields our most precise determination of the \ttbar cross section in a
single channel.
In the data selection we require the presence of one lepton ($e$ or $\mu$), a
large value of $\ebar_T$ ($\ebar_T>20$~GeV), and at least 3 jets with
$E_T>15$~GeV and $|\eta|<2.0$ ~\cite{eta}. 
The dominant backgrounds come from $W+$jets production and QCD multi--jet processes.
The selection criteria are identical to what is described in
Ref.~\cite{top_observation}, with a few exceptions.

Here, we outline the improvement used in this measurement and the differences with
the previous result.
The main difference in our measurement consists of a revised determination of
the efficiency of the SVX tagging algorithm. We measure the tagging efficiency
in the data and in simulated events.

The previous result used a method which determined the $b$--tagging
efficiency using a Monte Carlo (MC) simulation, tuned to give the same track finding	
efficiency as in the data.
The new method directly compares the $b$--tagging efficiencies in low--P$_T$
inclusive electron samples (rich in heavy flavor) from data and from simulation.
We find that there is no E$_T$ dependence of the data--to--MC $b$--tagging efficiency ratio,
alleviating previous concerns about extrapolating the results from low E$_T$
jets in the inclusive electron samples to jets in top events.
%The new method was checked using a larger MC sample than previously used.

\begin{figure}[thb]
\begin{minipage}[]{.46\linewidth}
\hspace*{-0.5cm}\centerline{\psfig{figure=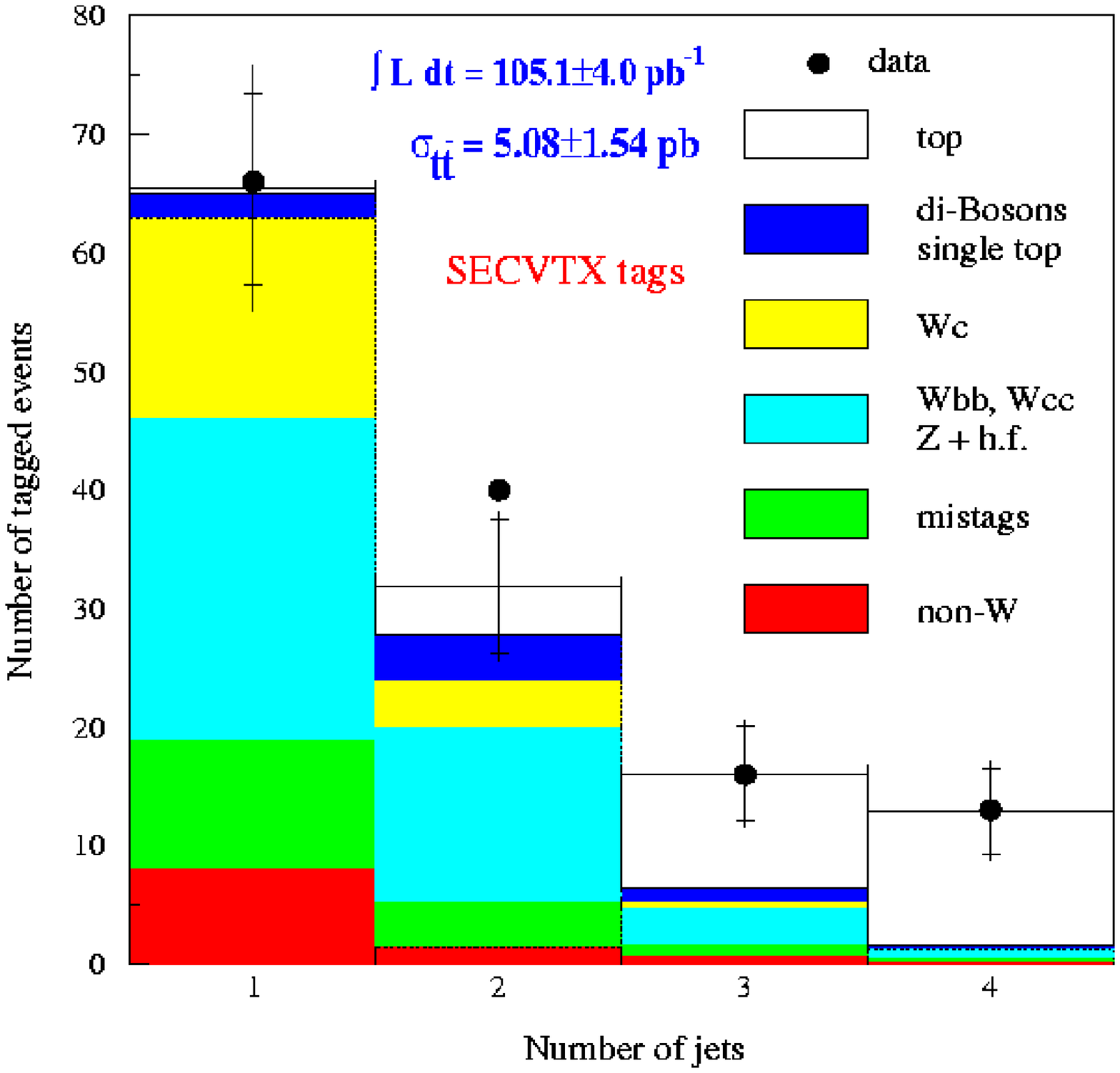,height=8.5cm,width=8.5cm}}
\caption{\label{W_plus_njets} 
Observed and expected number of events with SVX tags. The vertical bars
represent the overall uncertainty on the expected number of tags, and the
horizontal ticks on the bars mark the contribution from the statistical uncertainty.}
\end{minipage}
\hspace*{0.5cm}
\begin{minipage}[]{.46\linewidth}
\centerline{\psfig{figure=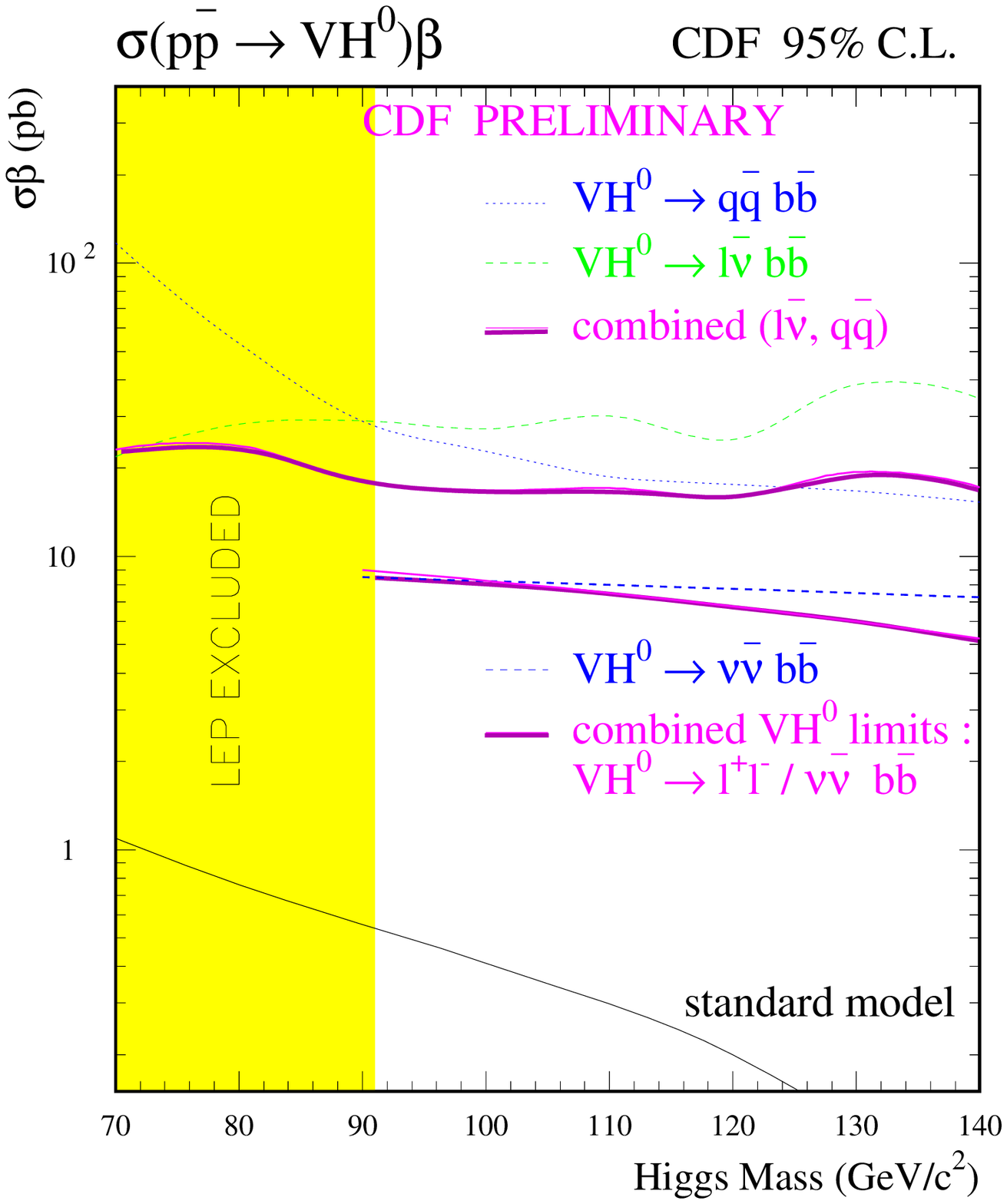,height=8.5cm,width=8.5cm}}
\caption{\label{higgs_limits} 
Upper limit on $\sigma(VH^{\circ})\cdot BR(H^{\circ}\rightarrow b\bar{b})$, 
where $V=W,Z$, from individual and combined 
$VH^{\circ}\rightarrow l\nu b\bar{b}, q\bar{q} b\bar{b}, l^+l^- b\bar{b}, \nu\bar{\nu} b\bar{b}$ channels.}
\end{minipage}
\end{figure}

Also, the Run I SVX $b$--tagging efficiency for a \ttbar event has increased
from $\approx$~43\% to $\approx$~50\%.
Most of the difference is due to outdated $b$--lifetimes
previously used in the MC, outdated $B$ and $D$ branching ratios, and slight
inconsistencies in the modeling of the SVX geometry.
Among other improvements, we use a slightly different event selection where we 
measure jet energies and $\ebar_T$
with the z--vertex of the primary lepton, instead of the center of the detector. 
Also, we have a better selection of
candidates from $Z^\circ$ bosons. The new selection finds about 20\% more real $Z^\circ$'s and rejects
about 20\% less fake $Z^\circ$'s.
We use a new method to calculate mistags that uses a parameterization of the
negative tagging rate from all jet data after corrections for heavy flavor
contribution to the negative tags. The associated uncertainty is 10\% (instead
of 40\%).
The rate of $W+\ge 1$~jets events is shown in Figure~\ref{W_plus_njets}.
We compute the \ttbar production cross section from the excess of events with at least one
$b$--tag in the $W+\ge 3$~jets sample. The whole excess is attributed to \ttbar events.
In 105.1$\pm$4.0~pb$^{-1}$ of data we find 29 $W+\ge$~3 jet events with an expected background
contribution of 8.0$\pm$1.0 events.
The resulting new value of the top cross section in the lepton+jets channel
using the SVX tagging algorithm is 5.1$\pm$1.5 pb.

We use the new value for the $b$--tagging efficiency, and apply the needed
modifications to the calculation of the cross section values in the other decay
channels. Then, we combine the results from the
``lepton+jets'' channel (including SLT) with
the updated results from the dilepton and from the all--hadronic
channels. We use a likelihood procedure, where we take into account the
correlations between the backgrounds.
The new result yields a combined \ttbar production cross section of
$\sigma_{t\overline{t}}=6.5^{+1.7}_{-1.4}$~pb, which can be compared to our
previous value of $7.6^{+1.8}_{-1.5}$~pb.

\section{Higgs Searches}

\subsection{SM Higgs}
The SM provides a simple mechanism for spontaneous symmetry breaking
through the introduction of a scalar field doublet. The doublet has one single
observable scalar particle, the Higgs boson, with an unknown mass but with fixed
couplings to other particles.
At the Tevatron one is more likely to observe the WH and ZH production processes.
In fact, although the Higgs production cross section from gluon--gluon fusion
($gg\rightarrow H$) is larger
(by a factor of $\sim 3$ for a Higgs mass of 100 GeV/c$^2$ ~\cite{Stange_Marciano}) 
than the associated production of 
$VH$ ($V=W^{\pm},Z^{\circ}$), the decay products from $H\rightarrow b\bar{b}, \tau\bar{\tau}$
are difficult to disentangle from the di--jet QCD background.
Instead, when the Higgs is produced in association with a $V$ boson, 
we can use the presence of an additional lepton or additional jets (possibly
from $W^{\pm}$ or $Z^\circ$) to discriminate 
from background sources and use the $b$--tagging 
algorithm to identify $b$--jets from $H\rightarrow b\bar{b}$ events.
The Tevatron is mostly sensitive in the region for 
$80 \le M_{H_{SM}}\le 130$ GeV/c$^2$~\cite{Gunion_Stange}, where the branching
ratio of ${\cal B} (H \rightarrow b\overline{b})$ is dominant.

We study various channels with different signatures in the final state.
We search for the SM Higgs in the lepton+jets channel, in $q\bar{q}\rightarrow WH\rightarrow 
		l\bar{\nu}b\bar{b},ZH\rightarrow llb\bar{b}$ events.
We look for an isolated high--$p_T$ lepton $+\ebar_T$ to identify $W$ decays or
pairs of leptons to identify $Z^{\circ}$ decays, and jets with $b$ tags to 
identify the $H\rightarrow b\bar{b}$ decays.
We also search for the Higgs in multi-jet events with $\ge 4$ jets
($q\bar{q}\rightarrow W/Z H\rightarrow jj b\bar{b}$), of which at least two
jets are $b$--tagged.
Finally, we look for signatures with large $\ebar_T$ from the process 
$q\bar{q}\rightarrow ZH\rightarrow \nu\bar{\nu}b\bar{b}$. In this analysis we
start with a $\ebar_T$ data sample (with $\ebar_T > 35$~GeV) and require at least 2 $b$--tagged jets.
We observe no deviations from the known SM backgrounds, in any of these final states.
Therefore, we combine the results from the different channels and obtain an upper 
limit on Higgs production. Figure~\ref{higgs_limits} shows the 95\% C.L. 
upper limit results on
$\sigma(VH^{\circ})\cdot BR(H^{\circ}\rightarrow b\bar{b})$.
The results indicate that the measurements at the Tevatron are not yet
sensitive to the discovery of the SM Higgs,
due to the small number of expected events.

\begin{figure}[thb]
\begin{minipage}[]{.46\linewidth}
\centerline{\psfig{figure=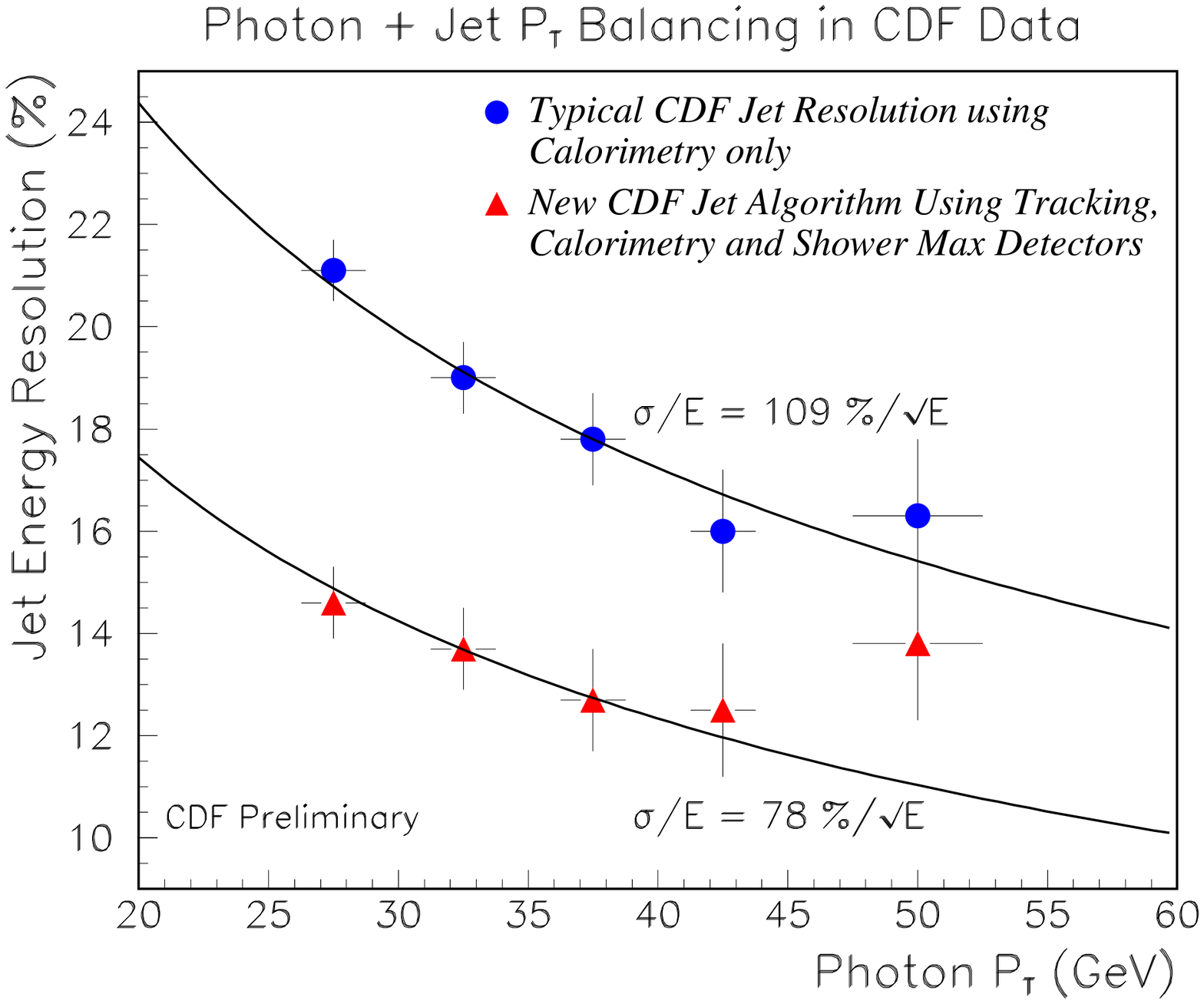,height=8.5cm,width=8.5cm}}
\caption{\label{jet_resolution} 
The jet energy resolution plotted as a function of photon $p_T$.
We compare the results from a new improved algorithm to our standard method.}
\end{minipage}
\hspace*{0.5cm}
\begin{minipage}[]{.46\linewidth}
\centerline{\psfig{figure=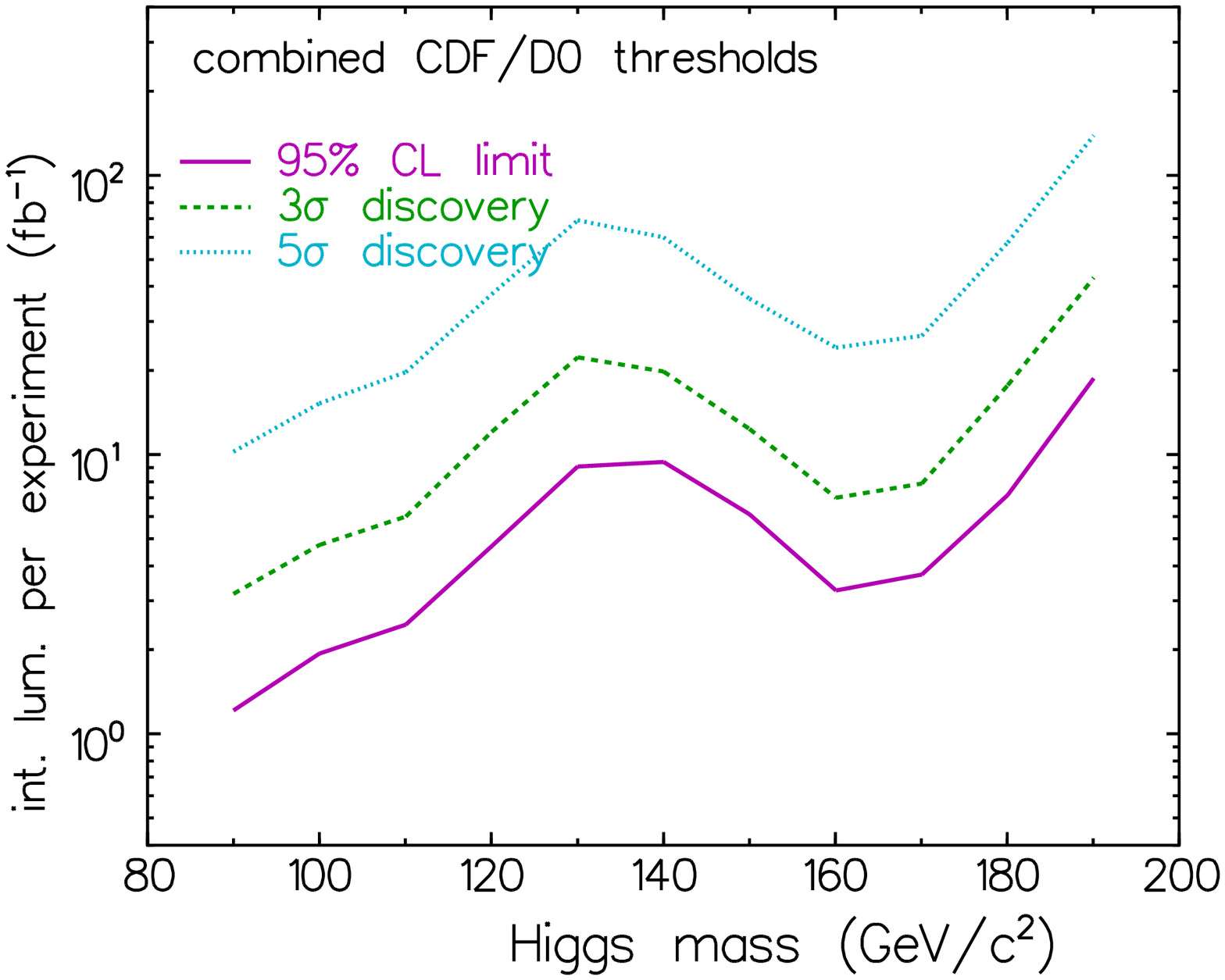,height=8.5cm,width=8.5cm}}
\vspace*{-4cm}
\caption{\label{higgs_run2} 
The integrated luminosity per experiment needed for a 5$\sigma$ and a 3$\sigma$
discovery, and 95$\%$ C.L. limit is plotted as a function of the Higgs mass.
CDF and D\O ~ results are combined as well as several channels. 
We include in the simulation an improvement of 30$\%$ in the di--jet
mass resolution over Run I, and the new Run II detector acceptance.}
\end{minipage}
\end{figure}

We have studied the feasibility of SM Higgs search in Run II.
We expect to improve our $b$--tagging capabilities, due to an
increased acceptance~\cite{conway} of our micro-vertex detector, by about 50\% in
events with single $b$ tagging.
Also in the future, the improvement of the di--jet mass resolution will 
be crucial for observing the Higgs boson. 
Therefore, we have studied a new algorithm for improving the jet 
energy resolution in the Run I data sample. We have measured 
an improvement of more than 25\%
as compared to our standard algorithm for single jet energy resolution, in a
photon+jet sample (see Figure~\ref{jet_resolution}). 
In di--jet events we could expect an improvement on the di--jet mass 
resolution of an additional factor of $\sqrt{2}$.
We combine results from the CDF and D\O ~ experiments at the Tevatron
as well as several channels. We include $WH$ and $ZH$ channels with the
Higgs decaying into $b\bar{b}$ pairs for a Higgs mass below 130 GeV, while for
a higher mass the $H\rightarrow WW$ decay has also been considered.
Figure~\ref{higgs_run2} shows the expectations once we include
a better di--$b$~jet mass resolution and the new detector acceptance.
The integrated luminosity needed per experiment for a 5$\sigma$ and 3$\sigma$
discovery, and 95$\%$ C.L. limit is plotted as a function of the SM Higgs mass.
Instead, if we use the present di--jet mass resolution, the luminosity
needed to achieve the same results approximately doubles.

\subsection{MSSM Higgs}
The Minimal Supersymmetric Model (MSSM)~\cite{MSSM} is a further 
extension of the SM.
In the MSSM, a more complex symmetry breaking mechanism occurs. Five observable
scalar particles are predicted: two charged ($H^{\pm}$) and three
neutral ($h^\circ, H^\circ$, and $A^\circ$) Higgs bosons.
%, and one CP--odd pseudo--scalar ($A^\circ$).
In the enlarged Higgs sector two new parameters are introduced: $\tan\beta$ which
is the ratio of the vacuum expectation values of the two Higgs doublets, and
the Higgs mass parameter $\mu$.

At the Tevatron, $h^\circ, H^\circ$ and $A^\circ$, can be produced similarly to
SM Higgs.
In the MSSM, at large values of $\tan\beta$, the Higgs couplings to $\tau^+\tau^-$ and 
$b\overline{b}$ pairs are greatly enhanced. The coupling is approximately
proportional to ~ $(\tan\beta)^2 \sim (m_{top}/m_b) \sim 1000$. This results in a
large production cross section in $p\overline{p}\rightarrow
b\overline{b}h^\circ, b\overline{b}H^\circ, b\overline{b}A^\circ$ events (the same
would happen for production in association with $\tau^+\tau^-$ pairs).
At CDF we have searched for events with $\ge 4$ jets (of which $\ge
2$~$b$--tagged jets) in the final state.
The number of events found in the data is in good agreement with the expected
backgrounds (mainly from QCD events, from mistags, and from $Z^\circ jj$). 
We can set a 95\%~C.L.
limit on the mass of MSSM neutral Higgs as a function of $\tan\beta$.
These results are preliminary and not presented at the time of this conference.

In the charged Higgs sector instead,
we have sensitivity at low values of $\tan\beta$ where $H^+\rightarrow c\bar{s}$,
and at large values of $\tan\beta$ where $H^+\rightarrow\tau\nu$.
We presented results on direct and indirect Higgs production from top
quark decays $t\rightarrow H^+b$, in both regions of $\tan\beta$.
We do not find any anomalous excess of events over the expected background and
set a limit on Higgs production. Results have not changed since ~\cite{michele}.\\
At large values of $\tan\beta$,
we can also use the results from a search for top quark pair production
in the e$\tau + \ebar_T$+jets and $\mu\tau +\ebar_T$+jets signatures~\cite{tau_dilepton}.
Here, we exploit the competing decays 
$t\rightarrow H^+b$ and $t\rightarrow W^+b$, when $W^+,H^+ \rightarrow \tau^+\nu$.
In the region at
large values of $\tan\beta$ the $tbH^+$ Yukawa coupling may become
non--perturbative (see  Ref.~\cite{Coarasa_Guasch_Sola}). Therefore we choose
to use a parameter--independent space to determine the excluded region.
We can set a 95\% C.L. upper limit on
the branching ratio of  ${\cal B} (t \rightarrow H^{+} b)$
in the range 0.5 to 0.6 at 95\% C.L. for Higgs masses in the range 60 to
160 GeV~\cite{prd_higgs}, assuming the branching ratio
for \mbox{$H^+\rightarrow \tau \nu$} is 100\% . 
The $\tau$ lepton is detected through its 1--prong and 3--prong hadronic decays.

\section{SUSY Searches: stop and sbottom quarks}

The range and complexity of SUSY signatures is growing fast. In the analysis of
Run I data it has become evident that the use of heavy flavor tagging is gaining
importance and it has become crucial. Here, we present the results from a
recent analysis which searches for scalar quarks in events with $\ebar_T $+jets.
The signature comes from the decays of the supersymmetric particles decaying to jets and to neutralinos.
We look for scalar top ($\tilde{t}_1$) and for scalar bottom ($\tilde{b}_1$) quarks
in events with similar final states~\cite{holck}.

\begin{figure}[htp]
\begin{minipage}[]{.46\linewidth}
\epsfysize=8.0cm
\epsfxsize=10.0cm
\epsfxsize=1.0\textwidth
\epsfbox[20 150 550 680]{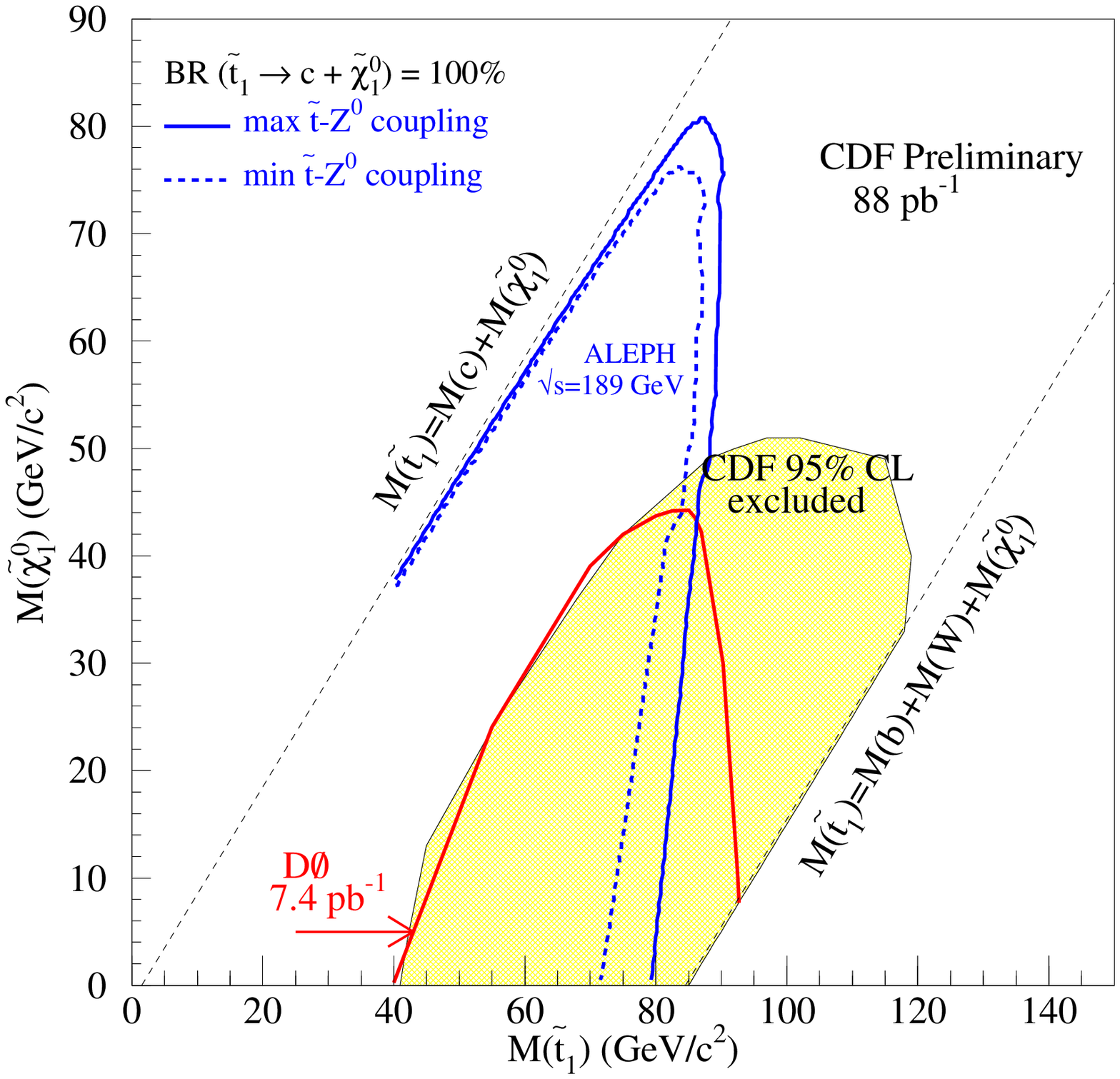}
\vspace*{-0.3cm}
\caption{
The 95\% C.L. exclusion region (shaded region) in the plane
$m_{\tilde{\chi^{\circ}_1}}$ versus $m_{\tilde{t}_1}$
for $\tilde{t}_1 \rightarrow c\chi^{\circ}_1$. Also shown are results from D\O ~and OPAL.}
\label{stop_limits}
\end{minipage}
\hspace{0.5cm}
\begin{minipage}[]{.46\linewidth}
\epsfysize=9.0cm
\epsfxsize=10.0cm
\epsfxsize=1.0\textwidth
\epsfbox[20 150 550 680]{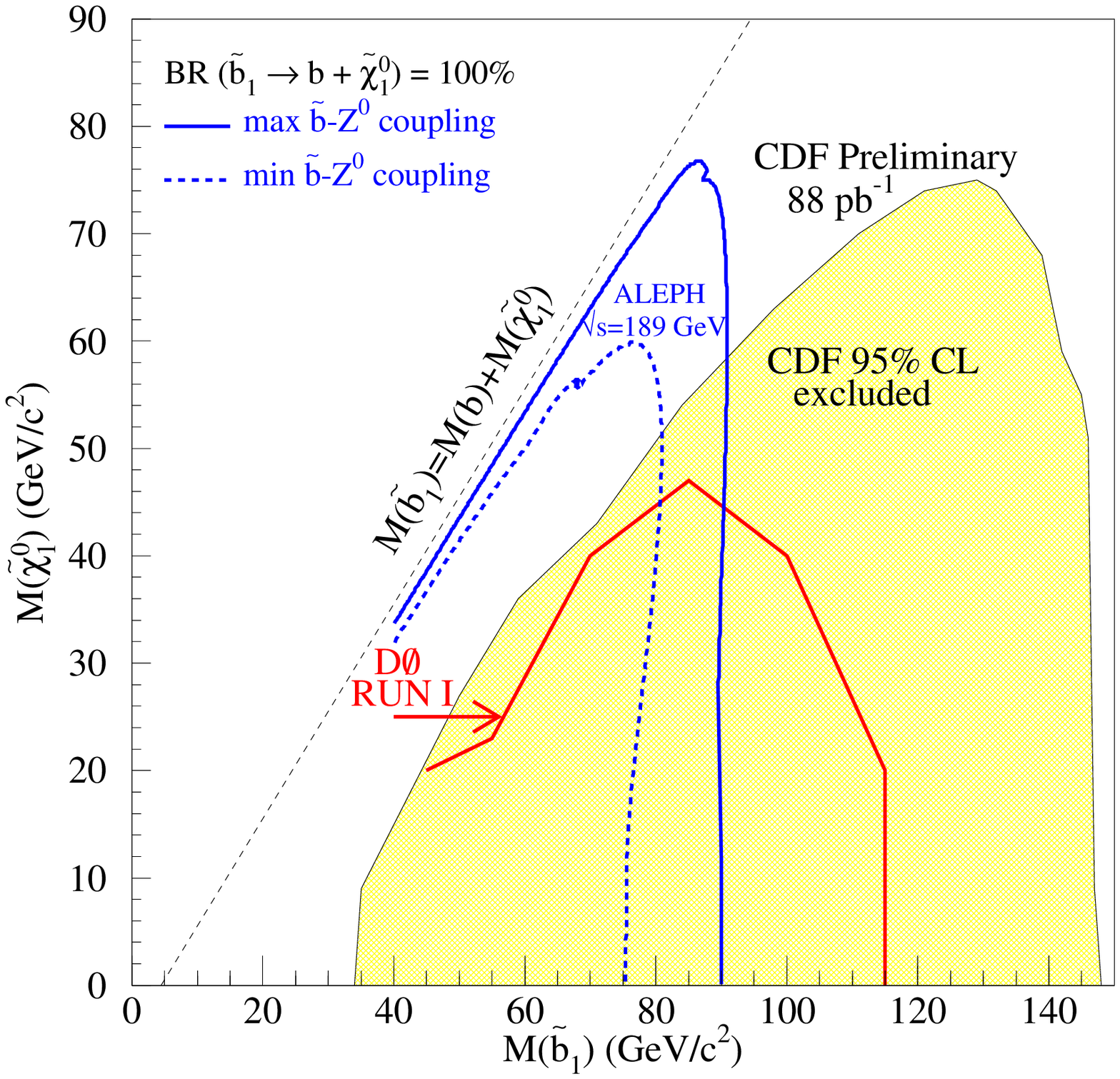}
\vspace*{-0.3cm}
\caption{The 95\% C.L. exclusion region (shaded region) in the plane
$m_{\tilde{\chi^{\circ}_1}}$ versus $m_{\tilde{b}_1}$
for $\tilde{b}_1 \rightarrow b\chi^{\circ}_1$. Also shown are results from D\O ~and OPAL.}
\label{sbottom_limits}
\end{minipage}
\end{figure}

We search for the scalar top quark in events with large $\ebar_T$ and with a pair
of heavy flavor jets, from the decay
$p\overline{p}\rightarrow \tilde{t}_1\overline{\tilde{t}_1}\rightarrow 
(c\tilde{\chi}^{\circ}_1)(\bar{c}\tilde{\chi}^{\circ}_1)$.
We assume that the neutralino $\tilde{\chi}^{\circ}_1$ is the lightest
supersymmetric particle and it is stable.
From a data sample of 88~pb$^{-1}$, we select events with two or three jets, 
no leptons, and with $\ebar_T >40$~GeV.
The requirement of large $\ebar_T$ eliminates almost completely the QCD di--jet background.
We require that $\ebar_T$ is neither parallel nor anti-parallel to any jet
direction in order to suppress contributions from QCD events where
missing energy comes from jet energy mismeasurement. 
We use a ``Jet Probability'' (JP) algorithm to tag jets with a heavy flavor content.
The JP algorithm
relies on determining a probability that a jet originates from a secondary
vertex, and it is optimized for charm or bottom selection in stop 
or sbottom searches. 
The probability that the tracks in a jet come from the primary vertex is flat
and distributed between 0 and 1 for tracks originating from the primary
vertex, and peaks at zero for tracks from a secondary vertex.
The efficiency of the JP algorithm to tag charm quarks
is checked in a charm enriched sample, with $D^*$ and $D^\circ$ mesons.
In our final data sample we find 11 events consistent with an expected background of
14.5$\pm$4.2 events from SM sources.
We interpret the null result in the scalar top search as an excluded region in
$m_{\tilde{t}_1}$ versus $m_{\tilde{\chi}^{\circ}_1}$ parameter space using a
background subtraction method ~\cite{caso}.
The 95\% C.L. limit excluded region is shown in Figure~\ref{stop_limits}.

Also, as we mentioned earlier, we can use similar selection criteria and 
search for scalar bottom quarks in the decay channel
$p\overline{p}\rightarrow \tilde{b}_1\overline{\tilde{b}_1}\rightarrow
(b\tilde{\chi}^{\circ}_1)(\bar{b}\tilde{\chi}^{\circ}_1)$.
After we apply our final selection cuts, we find 5
events in the data sample, with an expected background from SM sources of 5.8$\pm$1.8 events. 
Figure~\ref{sbottom_limits} shows the 95\% C.L. limit in the $m_{\tilde{b}_1}$
versus $m_{\tilde{\chi}^{\circ}_1}$ plane. 

\section{Conclusions and Prospects}

\begin{figure}[htp]
\begin{minipage}[]{.46\linewidth}
\epsfxsize=1.2\textwidth
\hspace*{-1cm}\epsfbox[20 150 550 680]{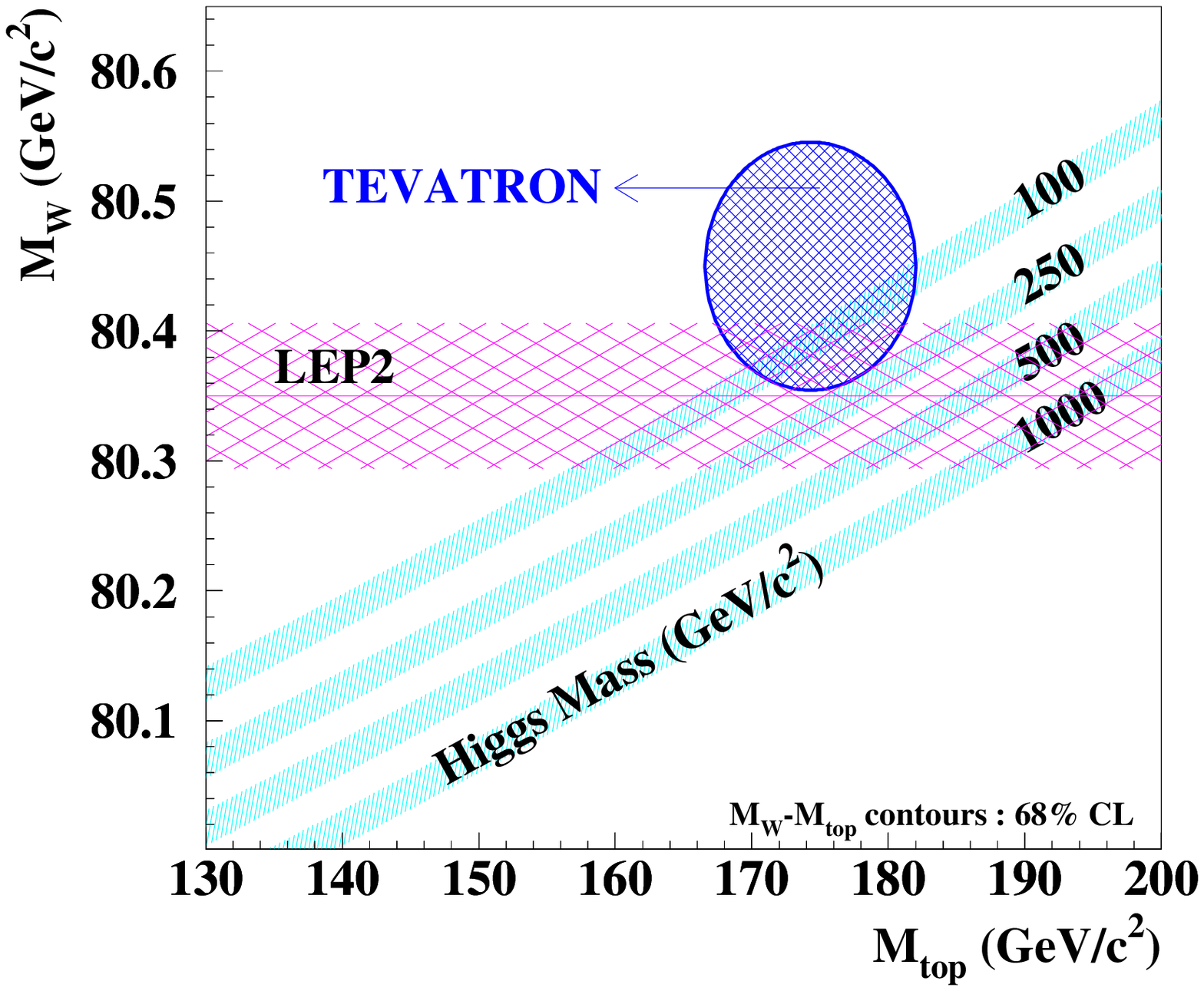}
\vspace*{-1.5cm}
\caption{
The measurements of the top quark and $W$ boson masses performed at the
Tevatron (from both CDF and D\O~) and at LEP2 
superimpose on different predictions for the value of the Higgs mass.}
\label{sm_higgs_run2}
\end{minipage}
\end{figure}

We have presented some of the latest results obtained at CDF during Run I. Still,
four years after the end of Run I, many analyses are being completed and many
are still in progress.
In Run II, both the CDF and the D\O~ detectors will be much more powerful, 
with much improved vertex detectors and
tracking system which allow for larger acceptance and for higher efficiency 
in flavor tagging.
The new upgraded detectors will collect more data at a higher energy of $\sqrt{s}=2$~TeV.
At this energy, the production cross section of heavy particles in the SUSY 
zoo for example, will increase significantly.
Furthermore, with a factor of 20 in expected delivered luminosity during Run II, the
experiments at the Tevatron will have a great chance to shed light into the new
world of ``new physics''.
Most importantly, by extending Run II up to an integrated luminosity of about 20~fb$^{-1}$
and combining search channels, the Tevatron can perform a crucial test of the
MSSM Higgs boson sector. The $W$ and top quark mass measurements,
from the Tevatron (both from CDF and from D\O~) and from the LEP2 experiments,
suggest a small value for the Higgs boson mass (see Figure~\ref{sm_higgs_run2}).
Also, most theories predict a Higgs mass which is comprised
between 130 and 150 GeV/c$^2$. When combining these informations together, the
picture looks promising for the Higgs searches at the Tevatron during Run II.
The experience gained from Run I analyses will greatly increase the quality of
the Run II searches. New triggering capabilities will open previously
inaccessible channels. In conclusion, the upcoming Run II at the Tevatron 
appears as an exciting field of promises.

\section{Acknowledgments}

I would like to thank the organizers of this conference for a wonderful
workshop and a marvelous hospitality.
In particular I appreciated the discussions and direct contact with
many of the participants at the conference to understand and update my
knowledge of the results from the current experiments, in the sparse field of
high--energy physics.
Most of all, I would like to thank my collaborators for providing me with all of these
results. In particular, I would like to thank Gerry Bauer for meticulously revising the
manuscript and for his many suggestions in the difficult path of life.
I would also like to thank Luc Demortier for his comments on the text
and Stefano Lami for his constant wisdom.

%            REFERENCES

%Tables and figures with captions must be build into the text or
% placed after the bibliography. Can be used \psfig or \epsfig .

%                 Fig.1/2 (B/W)

%                 Fig.3/4 (Color)

%                 Fig.5/6 (Color)

%                 Fig.7/8 (Color)

%                 Fig.9 (Color)


\begin{thebibliography}{99}
\bibitem{eta} In the CDF coordinate system, $\theta$ and $\phi$ are the 
polar and azimuthal angles, respectively, with 
respect to the proton beam direction. The pseudorapidity $\eta$ is defined as
$-\ln\tan(\theta/2)$. The transverse momentum of a particle is $P_T = P
\sin\theta$. The analogous quantity using energies, defined as $E_T = E
\sin\theta$, is called transverse energy.
The missing transverse energy $E\!\!\!\!/_T$ is defined as $-\sum E_T^i\ 
\hat{r}_i$, where $\hat{r}_i$ are the unit vectors in the transverse plane 
pointing to the energy depositions in the calorimeter.
\end{thebibliography}
\end{document}